\documentclass[aip,preprint,showpacs]{revtex4}
\usepackage{amssymb}
\usepackage{graphicx}
\usepackage[T1]{fontenc}
\usepackage{array}
\usepackage[american]{babel}

\usepackage{color}

\graphicspath{{images/}}

\begin{document}

\title{Magnetism tailored by mechanical strain engineering in PrVO$_3$
  thin films} \author{D. Kumar,$^1$ A. David,$^1$ A. Fouchet,$^1$
  A. Pautrat,$^1$ J. Varignon,$^2$ C. U. Jung,$^3$ U. L$\ddot{u}$ders,$^1$
  B. Domeng$\grave{e}$s,$^1$ O. Copie,$^4$ P. Ghosez,$^2$ W. Prellier$^1$}
\thanks{wilfrid.prellier@ensicaen.fr} \affiliation{$^1$Laboratoire
  CRISMAT, CNRS UMR 6508, ENSICAEN, Normandie Universit\'e, 6 Bd
  Mar\'echal Juin, F-14050 Caen Cedex 4, France}
\affiliation{$^2$Physique Th\'eorique des Mat\'eriaux, Universit\'e de
  Li\'ege (B5), B-4000 Li\'ege, Belgium} \affiliation{$^3$Department
  of Physics, Hankuk University of Foreign Studies, Yongin, Gyeonggi
  17035, Korea} \affiliation{$^4$Institut Jean Lamour, CNRS-Universit\'e de Lorraine, Campus Artem, 2 all$\acute{e}$e Andr$\acute{e}$ Guinier BP 50840, 54011 Nancy cedex, France}

\begin{abstract} 
Transition-metal oxides with an \textit{ABO$_3$} perovskite structure
exhibit strongly entangled structural and electronic degrees of
freedom and thus, one expects to unveil exotic phases and properties
by acting on the lattice through various external stimuli. Using the
Jahn-Teller active praseodymium vanadate Pr$^{3+}$V$^{3+}$O$_3$
compound as a model system, we show that PrVO$_3$ N\'eel temperature
{\em T$_N$} can be raised by 40 K with respect to the bulk when grown as
thin films. Using advanced experimental techniques, this enhancement
is unambiguously ascribed to a tetragonality resulting from the
epitaxial compressive strain experienced by the films. {\em First-principles} simulations not only confirm experimental results,
but they also reveal that the strain promotes an unprecedented
orbital-ordering of the V$^{3+}$ $d$ electrons, strongly favoring
antiferromagnetic interactions. These results show that an accurate control of structural aspects is the key for
unveiling unexpected phases in oxides.  \quad
\end{abstract}
\maketitle
\newpage

\section{Introduction}

Transition metal oxides with an ABO$_3$ perovskite structure are
multi-functional materials displaying a large collection of properties
such as ferroelectricity, metal-to-insulator transition, high
\textit{T$_C$} superconductivity and colossal magneto-resistance (CMR)
for instance \cite{MIT,LMO,HSC,CMR}. This richness of physical behaviors
emerges through strongly coupled structural, electronic and magnetic
degrees of freedom, enabling possibilities to control the material's
properties with external stimuli \cite{Interface physics}. Among all approaches, strain
engineering allowed by minute deposition of oxides as thin films on a
range of commercially available substrates is likely the most adopted
strategy to unveil hidden phases in bulk. Most striking examples
achieved with strain engineering are (i) the observation of
ferroelectricity in SrTiO$_3$ films under tensile epitaxial strain
\cite{nature2004}, an otherwise quantum paralectric compound in bulk;
(ii) the rich ferroelectric phase diagram of BiFeO$_3$ as a function
of the applied epitaxial strain \cite{ReviewSandoBibes} or (iii) the
control of magneto-resistive properties in R$_{1-x}$A$_x$MnO$_3$ films
(R=rare-earth, A=Ca, Sr) \cite{RAMnO3,fontcuberta}.

In the search of multi-functional materials with possibly
unprecedented properties, one must consider materials with nearly
degenerate ground states that could be tailored by epitaxial
strain. Along with the widely studied rare-earth manganites \cite{RMOphase,RMOfe,structure et magnetic},
rare-earth vanadate perovskites \textit{R}VO$_3$ (\textit{R}=Lu-La, Y) are
prototypical compounds showing strongly coupled
structural-spin-orbital properties \cite{orbital-spin,charge}. At high
temperature, \textit{R}VO$_3$ compounds are paramagnetic insulators adopting
the usual orthorhombic $Pbnm$ symmetry displayed by perovskites and
characterized by $a^-a^-c^+$ octahedral rotations. Due to the
intrinsic instability displayed by the V$^{3+}$ $t_{2g}^2$ electronic
configuration \cite{VarignonZunger-originGapsABO3}, a Jahn-Teller
distortion appears and induces a symmetry lowering to a monoclinic
$P2_1/b$ structure at the temperature $T_{oo}$. It produces a G-type
orbital-ordering with alternating occupancy of the $d_{xz}$ and
$d_{yz}$ orbitals on neighboring V sites according to a rock-salt like
pattern -- the second electron is located in the low energy $d_{xy}$ orbital on
all V sites. It is then followed by a magnetic transition at
$T_N$<$T_{oo}$ -- except for LaVO$_3$ for which $T_N$ is 2 K above
$T_{oo}$ \cite{bulkPVO} -- to a C-type AFM order explained by
Kugel-Khomskii and Goodenough-Kanamori rules
\cite{Kugel-Khomskii,Goodenough}. Finally, for vanadates
involving rare-earth with a small ionic radius (R=Lu-Dy, Y), the
compound goes back to an orthorhombic $Pbnm$ symmetry characterized by
an alternative Jahn-Teller motion producing a C-type orbital
arrangement of $t_{2g}$ orbitals -- columnar arrangement along the $c$
axis of alternating $d_{xz}$ and $d_{yz}$ orbitals -- that is
associated with a G-type AFM order at $T_{N2}$.

It is obvious that the chemical pressure induced by A site cations
dramatically influences the electronic and magnetic states of the
vanadates. Likewise, external stimuli such as hydrostatic pressure or
partial A site substitution can also tune the material properties
\cite{superexchange,pressure controlled OO}. Regarding thin films, a precise control of
oxygen vacancies concentration in PrVO$_3$ grown on a SrTiO$_3$
substrate was recently shown to produce a substantial chemical strain,
offering a pathway to modify the N\'eel temperature on a range of 30 K
using a unique substrate type \cite{chemical strain engineering}. Nevertheless,
basic questions remain largely unexplored in these compounds: {\em can
  we tune the vanadate properties using various epitaxial strains, and
  eventually promote new electronic phases?}  Aiming at providing
answers to these important questions, we have studied the effect of
epitaxial strains on the praseodymium vanadate perovskite using
advanced experimental techniques. We show that the N\'eel temperature
can be continously raised by 40 K with respect to the bulk by
increasing the compressive epitaxial strain. Our {\em
  first-principles} simulations confirm the experimentally observed
trend for $T_N$, but amazingly, they also reveal that this strong
enhancement is associated with an unprecedented orbital-order of
$t_{2g}$ levels.

\section{Methods}
{\em Experiments:} PrVO$_3$ (PVO) thin films ({\em t} $\sim$ 50 nm)
were grown on various substrates such as (110)-YAlO$_3$ (YAO),
(100)-LaAlO$_3$ (LAO), (100)-(La,Sr)(Al,Ta)O$_3$ (LSAT), and
(100)-SrTiO$_3$ (STO) using the pulsed laser deposition (PLD)
method. A KrF excimer laser ($\lambda$ = 248 nm) with repetition rate
of 2 Hz and laser fluence of $\sim$ 2 \textit{J/cm$^2$} was focused on
stoichiometric ceramic targets. All the films used in this study were
deposited at an optimum growth temperature (\textit{T}$_G$) of 650
$^\circ$C and under oxygen partial pressure ($\textit{P}_{O_2}$) of
10$^{-6}$mbar. The thickness of PVO films was kept nearly constant at
50 nm. To identify the lattice mismatch, the pseudo-cubic lattice
parameters of YAO, LAO, LSAT and STO were used as: 3.700 {\AA}, 3.790
{\AA}, 3.868 {\AA} and 3.905 {\AA} respectively. The crystallinity and
the structure were characterized using conventional High resolution x-ray diffraction
(XRD) technique (Bruker D8 Discover diffractometer, Cu
\textit{K}$_{{\alpha}1}$ radiation, $\lambda$ = 1.54056 {\AA}). The
surface morphology was investigated using atomic force microscopy
(AFM) PicoSPM. The Resistivity ($\rho$(\textit{T})) measurements were
performed using the four point probe technique in a Quantum Design
Physical Properties Measurement System (PPMS). The magnetic
measurements were obtained using Superconducting Quantum Interface
Device magnetometer (SQUID), as a function of temperature (\textit{T})
and magnetic field (\textit{H}). Transmission Electron Microscopy
(TEM) - Scanning Transmission Electron Microscopy (STEM) study was
carried out on a JEM-ARM200F, operating at 200 kV, equipped with a cold
Field Emission Gun and double TEM-STEM Cs correctors, ensuring lattice
TEM or STEM image resolution below 0.1 nm, and JEOL EDS system. Thin
TEM lamellae were prepared in a Dual-Beam system (FEI-HELIOS 600)
equipped with Easy-lift manipulator designed for In-situ Lift-Out thin
lamella preparations.

{\em Theoretical calculations:} {\em first principles} calculations
are performed using Density Functional Theory with the VASP package
\cite{VASP1,VASP2}. We have employed the PBEsol functional in addition
to a U potential on V $d$ levels of 3.5 eV, entering as a single
effective parameter \cite{LDA-Dudarev}, in order to better cancel the
spurious self-interaction term. This parameter was fitted in
References \onlinecite{chemical strain engineering,coupling} and was providing
correct electronic, magnetic and structural features for PrVO$_3$
ground state. Pr $4f$ electrons are not considered in the study and
are included in the Projected Augmented Wave (PAW
\cite{PAW}potential. Unit cells used in our simulations correspond to
a (2a,2a,2a) cubic cell allowing for the oxygen cage rotations and
Jahn-Teller motions to develop ({\em i.e.} 8 formula units). The
energy cut-off is set to 500 eV and a $4\times4\times4$ kpoint mesh is
employed. Four magnetic states are explored in our simulations, namely
the C, G and A-type SO as well as a ferromagnetic solution. We have
considered two growth orientations for the films with the in-phase
rotation axis ({\em i.e} the (001) $Pbnm$ axis) lying either along the
substrate or perpendiculary to it. We then block two PrVO$_3$ lattice
parameters to those of the substrate and relax the magnitude of the
remaining lattice parameter, although restriting it to be orthogonal
to the substrate due to the presence of 90$^{\circ}$ oriented domains
\cite{chemical strain engineering}.

Nearest neighbor magnetic exchange integrals J$_1$ and J$_2$,
corresponding to interactions along the (110) (or (1-10)) and (001)
directions respectively, are extracted by mapping energies of FM, A, C
and G-type spin orderings on an Heisenberg model of the form
$\hat{H}=-J\sum_{i<j}S_i.S_j$ where the sum runs over all possible
sites i and j in the cell. The cell is fixed to the ground state
structure for each strain value. In order to avoid modifications of
the electronic structure due strongly entangled spin-orbital degrees
of freedom in PrVO$_3$ \cite{Kugel-Khomskii}, we have frozen the $d$
orbital occupancies to that of the lowest energy state using the
modified DFT+U routine of VASP \cite{Watson} and we simply
switched spin channels to account for the magnetic order. The N\'eel
temperature is then computed using a mean-field model with $T_N
\propto (2J_1 + J_2)$. \\

\section{Results and discussion} 

We have grown a series of PrVO$_3$ (PVO) thin films using Pulsed Laser
Deposition on (001) oriented substrates {\em a priori} yielding either
nearly no epitaxial strain (SrTiO$_3$ (STO) substrate) or compressive strain
((La,Sr)(Al,Ta)O$_3$ (LSAT), LaAlO$_3$ (LAO) and YAlO$_3$ (YAO) substrates) {\em with respect to} the bulk PVO (see Figure 1a).


\begin{figure}
\centering \includegraphics[width=0.9\textwidth]{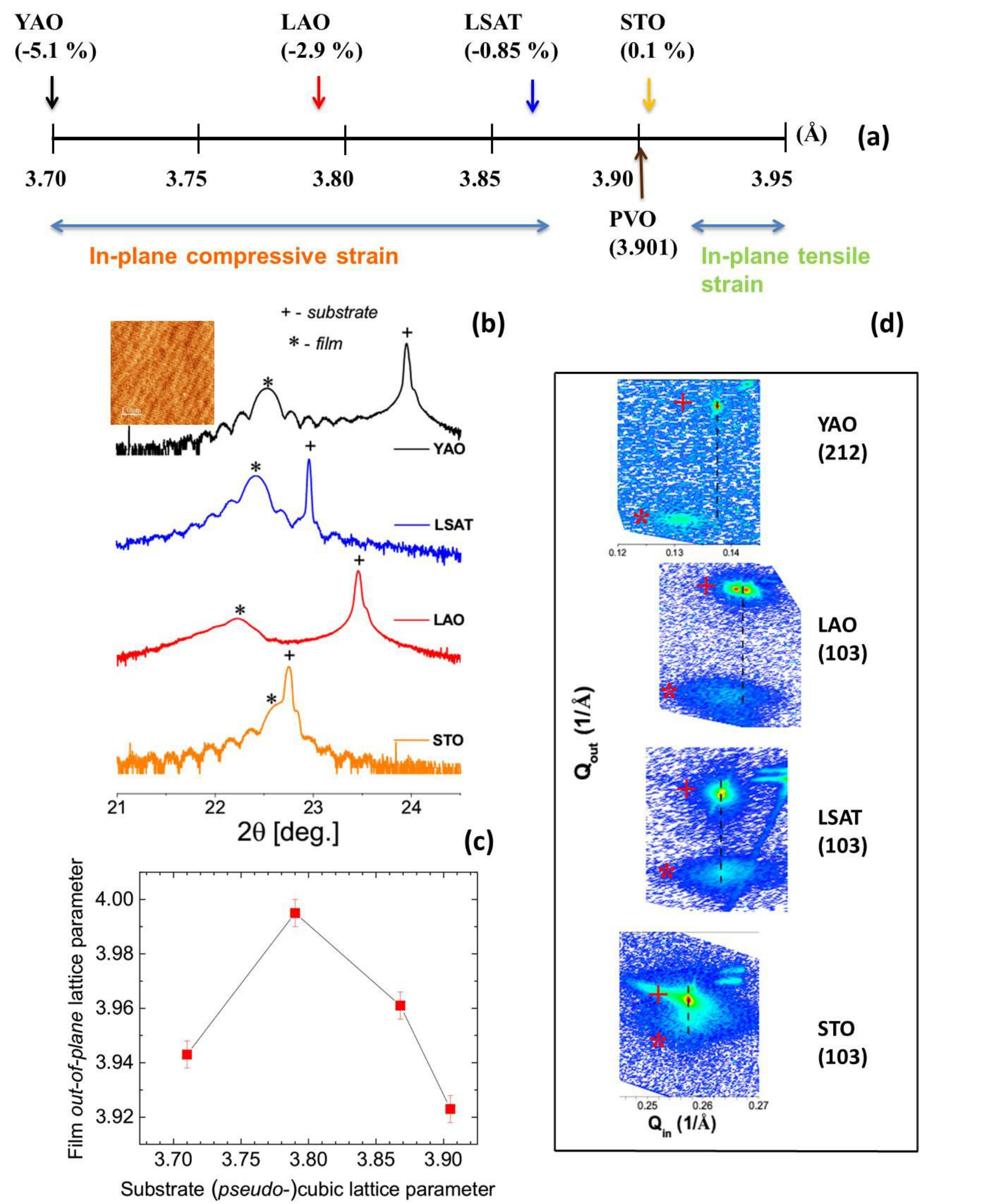}
\caption{(a) Lattice mismatch between bulk stoichiometric PrVO$_3$ and
  the various substrates at room temperature. (b) $\theta$-2$\theta$
  HRXRD scan for a series of 50 nm thick PVO thin films on different
  substrates. (001)$_C$ diffraction peak of the PVO film and substrate
  (LSAT, LAO, STO) and (110)$_c$ of YAO substrate are indicated by $\ast$
  and +, respectively. The inset of Figure 1b is a 5 ${\mu}$m X 5
  ${\mu}$m AFM image of PVO film deposited on LAO substrate (RMS
  surface roughness $\sim$ 2.3 \AA). (c) Evolution of the
  \textit{out-of-plane} lattice parameter as a function of the
  substrate pseudo-cubic lattice parameter. (d) RSMs around
  pseudo-cubic (103) plane of STO, LAO and LSAT, and (212) plane of
  YAO substrate. The substrate peaks are sharp, intense, shown by
  plus (+) sign and located on the upper region of RSM image. The
  film peaks, shown by asterisks ($\ast$) are broader and located on the lower region of the image. The solid and dotted lines are only guide to the eyes.}
  \label{fig-XRD}
\end{figure}

Figure 1b displays $\theta$ - 2$\theta$ scan for the epitaxially grown
PVO thin films. For most of the substrates, clear thickness fringes are observed around the
main diffraction peaks, confirming a uniform thickness and smooth interfaces of
the films. The film thickness estimated using these fringes in the diffraction pattern is actually around 50 nm for
all films leading to a growth rate ($\sim$ 0.09 {\AA}/\textit{pulse}).  In
the case of LAO substrate, these oscillations are however small and
subtle, probably due to presence of twin domains in the LAO substrate
\cite{CIO}. The films surfaces are quite smooth, presenting clear steps and terraces (see inset of Figure 1b). For example, the RMS surface
roughness of the PVO/LAO film was found around 2.3 {\AA}
indicating a flat surface. The evolution of the \textit{out-of-plane}
lattice parameter (calculated from XRD data) is plotted as a function of
substrate lattice parameter in Figure 1c. Surprisingly, it presents a maximum for LAO substrate and a relatively
lower lattice parameter for PVO/YAO film. This indicates the ability
of PVO/LAO film to adopt large strain and a lower or no strain in PVO/YAO film, which is anticipated for such a large
lattice mismatch.

To identify the strain states, reciprocal space maps were recorded
around (103)$_{\textit{c}}$ (where the index {\em c} refers to the cubic perovskite sublattice) planes of LAO, LSAT and STO and (212)
plane of YAO (Figure 1d). The X-ray reciprocal space
mapping shows well-developed film peaks in the lower
region and strong substrate peaks in the upper region for all the PVO
films. Since the horizontal peak
positions of the PVO film coincide with those of the substrate for both
LSAT and STO, we deduce that the film is fully strained with the
substrate, and has the same \textit{in-plane} lattice constant.
In the case of LAO, the small shift of the film peak to lower Q$_{in}$
value suggests an increase of the \textit{in-plane} lattice parameter,
and a partially relaxed film, which confirms a flexibility of the PVO structure for a large strain associated with large
lattice mismatch of -2.9 \%. Finally, we see that the PVO film is
fully relaxed on YAO, indicating that the growth is not coherent for
this peculiar substrate, which can be explained by large compressive
lattice mismatch (-5.1\%). Additionally, the film relaxes in order to minimize the accumulated strain
energy \cite{CIO,ferroelectricity,strain}.

The PVO unit-cell volume (pseudo-cubic) was extracted and reported in
table 1. The lattice volume is slightly reduced
compared to the bulk value (59.36 \AA$^3$) for PVO/YAO, PVO/LAO,
PVO/LSAT and increased for PVO/STO. This means that for tensile and
compressive strain, the conservation of volume is not perfect due to
non ideal Poisson's ratio \cite{CaMnO}.
Moreover, the \textit{out-of-plane} lattice parameter is well above
the bulk value (3.901 \AA) for all PVO films irrespective of the
strain (\textit{compressive/tensile}). This is due to low oxygen partial
pressures used during the growth which induces oxygen vacancies in the film, resulting in an enhancement of lattice parameter \cite{oxygen vacancies,PVO on STO}. Nevertheless, with increase of the
\textit{in-plane} compressive strain, the \textit{out-of plane}
lattice parameter is enhanced as expected when going from LSAT to LAO
substrate. The \textit{out-of-plane} lattice parameter of
PVO/YAO film is however much smaller, which is in agreement with a relaxed
film as shown in Figure 1d.
\begin{center}
\begin{table}
\caption{Summary of observed lattice parameters, experimental lattice
  mismatch, pseudo-cubic cell volumes, distortion (ratio of the
  \textit{out-of-plane} to \textit{in-plane} PVO lattice parameter),
  residual strain calculated {\em out-of-plane} i.e $\epsilon_{110}$, N\'eel temperature ({\em T$_N$}) and coercivity ({\em H$_c$}) of PVO films.}
\centering 
\begin{tabular}{ | m{5em} | m{1.5cm} | m{1.5cm} | m{1.5cm} | m{1.5cm} | m{1.5cm} | m{1.5cm} | m{1.5cm} | m{1.5cm} | m{1.5cm} | } \hline
  \textbf{Substrate}& \scriptsize \textbf{\textit{In-plane} lattice
    parameter of substrate [\AA]} &\scriptsize \textbf{Lattice
    mismatch (\%)}& \scriptsize \textbf{\textit{In-plane} lattice
    parameter of film [\AA]}& \scriptsize
  \textbf{\textit{Out-of-plane} lattice parameter of film [\AA]}&
  \scriptsize \textbf{Pseudo-cubic unit cell volume [\AA$^3$]}&
  \scriptsize \textbf{Distortion}& \scriptsize \textbf{Residual strain ({\em out-of-plane})
    (\%)}& \scriptsize \textbf{\textit{T$_N$} (K)}& \scriptsize
  \textbf{\textit{H$_c$} (T)} \\ \hline \hline \textbf{YAO} & 3.710 &
  -5.15 & 3.860 & 3.943 & 58.75 & 1.022 & 1.077 & 134 & 2.40 \\ \hline
  \textbf{LAO} & 3.790 & -2.93 & 3.830 & 3.995 & 58.60 & 1.043 & 2.412
  & 172 & 3.25 \\ \hline \textbf{LSAT} & 3.868 & -0.85 & 3.868 & 3.961
  & 59.26 & 1.024 & 1.540 & 125 & 2.70 \\ \hline \textbf{STO} & 3.905
  & 0.10 & 3.905 & 3.923 & 59.82 & 1.005 & 0.566 & 100 & 1.58
  \\ \hline
\end{tabular}
\end{table}
\end{center} 
The residual strain was calculated using : ${\epsilon_{hkl}}$ =
${(d^{0}_{hkl} - d_{hkl})/d^{0}_{hkl}}$ ; where $d^{0}_{hkl}$ and
$d_{hkl}$ are the pseudo-cubic PVO bulk and film {\em out-of-plane} lattice parameters
respectively. Interestingly, across the series, the measured strain increases from $\sim$ 0.5 \% for PVO/STO to $\sim$ 2.4 \% for PVO/LAO (table 1). Furthermore, albeit the relaxing behavior of PVO film on YAO substrate, the calculated strain is larger than PVO/STO (see Figure 1d for RSM and table 1 for strain values). This inconsistency could be explained as below.
As proposed by Herranz \cite{oxygen vacancies}, the STO substrate acts as oxygen reservoir during deposition and consequently the film behaves like a source of the oxygen vacancies. Therefore, oxygen vacancies tend to diffuse from film into the STO substrate, making film deficient of oxygen vacancies. As a consequence, the strain in PVO/STO film is as a result of lattice mismatch which is significantly small (0.1\%) and only partially due to oxygen vacancies. It is the other way round in PVO/YAO case {\em i.e.} the strain is induced by the oxygen vacancies in the film, and the impact of substrate in building the strain is minimum.
 
\begin{figure}[hbt!]
\centering
\includegraphics[width=0.8\textwidth]{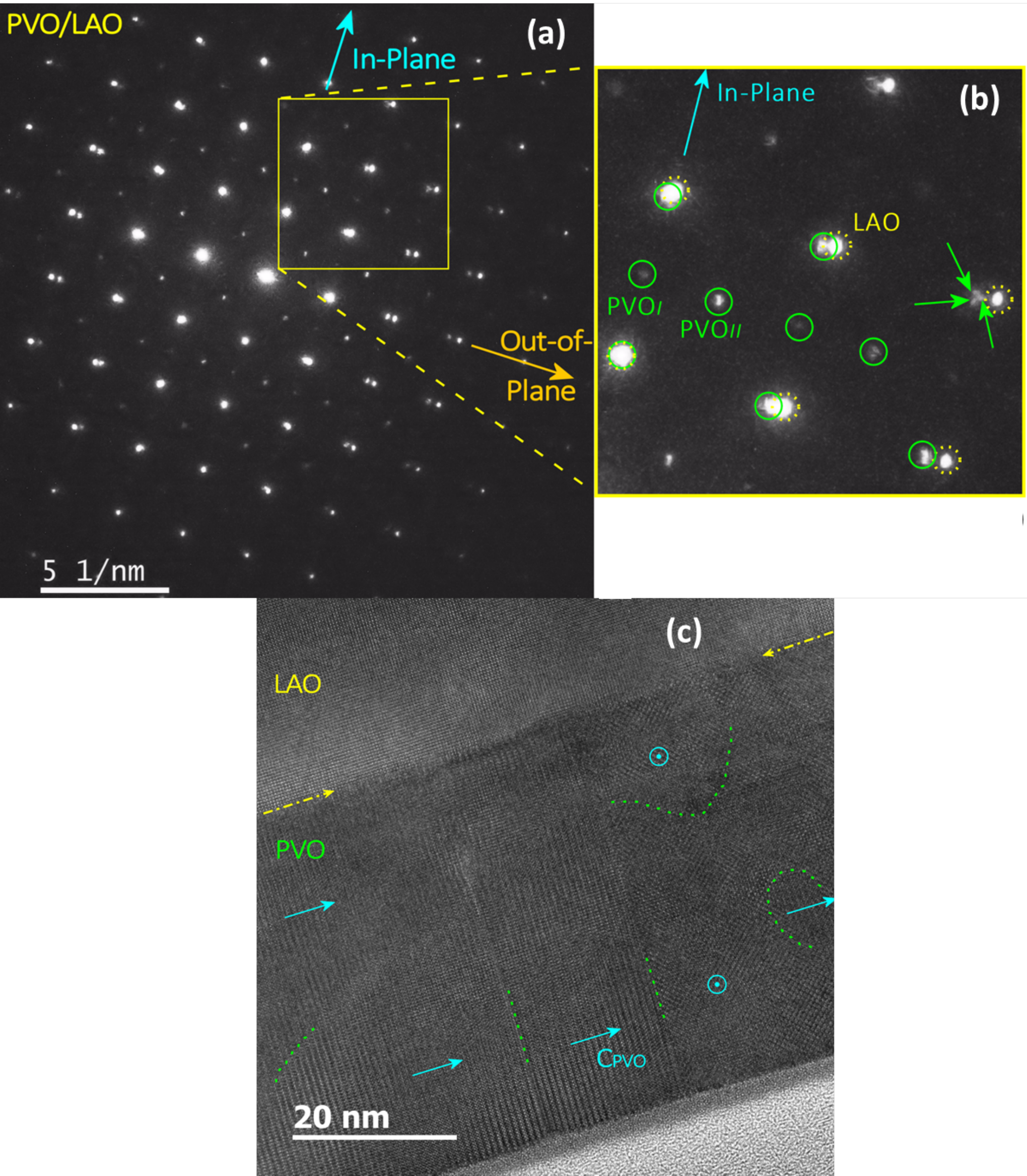}
\caption{(a) Typical SAED pattern of PVO/LAO sample; several patterns
  can be distinguished, especially along \textit{out-of-plane}
  direction. (b) Enlargment showing the complex splitting of PVO dots
  (green arrows). (c) HRTEM image of PVO film on LAO substrate. The orientation of
  c$_{PVO}$ is given on several domains. It always lies
  \textit{in-plane}, either parallely (cyan solid arrows) or perpendicularly
  (dotted circle) to the lamella plane. Domains are outlined with
  dotted lines. The interface between LAO substrate and PVO film
  exhibits contrast perturbation indicative of strains, depicted by dashed arrows.}
\end{figure}

In order to obtain details of the microstructure, Transmission
Electron Microscopy (TEM) studies were performed on cross-sectionnal
thin lamellae prepared for each sample. The lamellae were oriented in order to observe both the \textit{out-of-plane} axis,
i.e. growth direction, and one \textit{in-plane} axis, characteristic of the perovskite structure. The TEM study, through Electron
Diffraction (ED), High Resolution TEM and Scanning-TEM imaging allowed
a local characterization of the PVO films, in terms of orientation
with respect to the substrate, evolution of the parameters (strain),
nanostructure (domains) and quality of the film-substrate interface. A summary of the main observation is given in Table 1 of supporting information and more details can be found elsewhere
\cite{preparation}. The observed thickness of the PVO films is close to those calculated from XRD around 50 nm. The Selected Area Electron
Diffraction (SAED) study is in complete agreement with X-ray Reciprocal Space
Mapping. Almost no strain is observed on STO subtrate with a perfect
adequation of \textit{in} and \textit{out-of-plane} lattice
parameters (deduced from a perfect superposition of diffraction
spots of substrate and film). In the case of YAO substrate, two electron diffraction patterns can be clearly distinguished, one exhibiting YAO parameters and the second related to PVO parameters, along both {\em in} and {\em out-of-plane} directions (figure 2 in supporting information). Thus, there is almost no interaction between YAO substrate and PVO film.

On LSAT and LAO substrates, both parameters are influenced : the
strain being compressive, the PVO \textit{in-plane} lattice parameter
is decreased to fit the one of the substrate, leading to an increase of the
\textit{out-of-plane} lattice parameter. The PVO films always exhibit
small domains (several tens of nanometers)(table 1 of supporting information). In most of the observations, the PVO [001]$_o$ lies {\em in-plane}, and the
diffraction spots related to 2 x a$_{pc}$ along growth direction are
either weak (STO, YAO) or nonexistent (LAO).

The SAED pattern shown in Figure 2a,b illustrates these observations for PVO film grown on LAO substrate: several patterns are
superimposed, one LAO and two PVO ones. The latter correspond to
several diffracting domains (labelled \textit{I} and \textit{II} in Figure 2b) having the [110] reciprocal axis {\em out-of
plane}. Moreover, the enlargment of SAED pattern shows a more complex
splitting of PVO dots, that could be due to deformation of PVO
framework from one domain to the other. The domain size was evaluated
from several TEM images, covering about 0.5 ${\mu}$m of the PVO
film. It appeared that despite an apparent columnar growth, several
domains may be observed from the bottom to the surface of the film
(Figure 2c). In addition, measurements suggest that domains are smaller in size
when the PVO film is not strained (on STO and YAO substrates). Stacking
faults were observed in the upper part of the PVO film, on about 1/4
of the thickness and usually extend parallely to the growth
direction. They involve either the oxygen framework or both oxygen and cation ones.
  
To investigate the effect of biaxial strain on the physical
properties, the transport properties ($\rho$(\textit{T})) of PVO
films were investigated (see section 4 of supporting information). The insulator-like
$\rho$(\textit{T}) behavior was observed for PVO films on LAO, LSAT and
YAO. On the contrary, the PVO/STO film displayed a conducting-like
behavior, which is likely resulting from the presence of the oxygen
vacancies in STO substrate \cite{PVO on STO,oxygen vacancies,Hall mobility,two components}.

\begin{figure}[hbt!]
\centering
\includegraphics[width=1.0\textwidth]{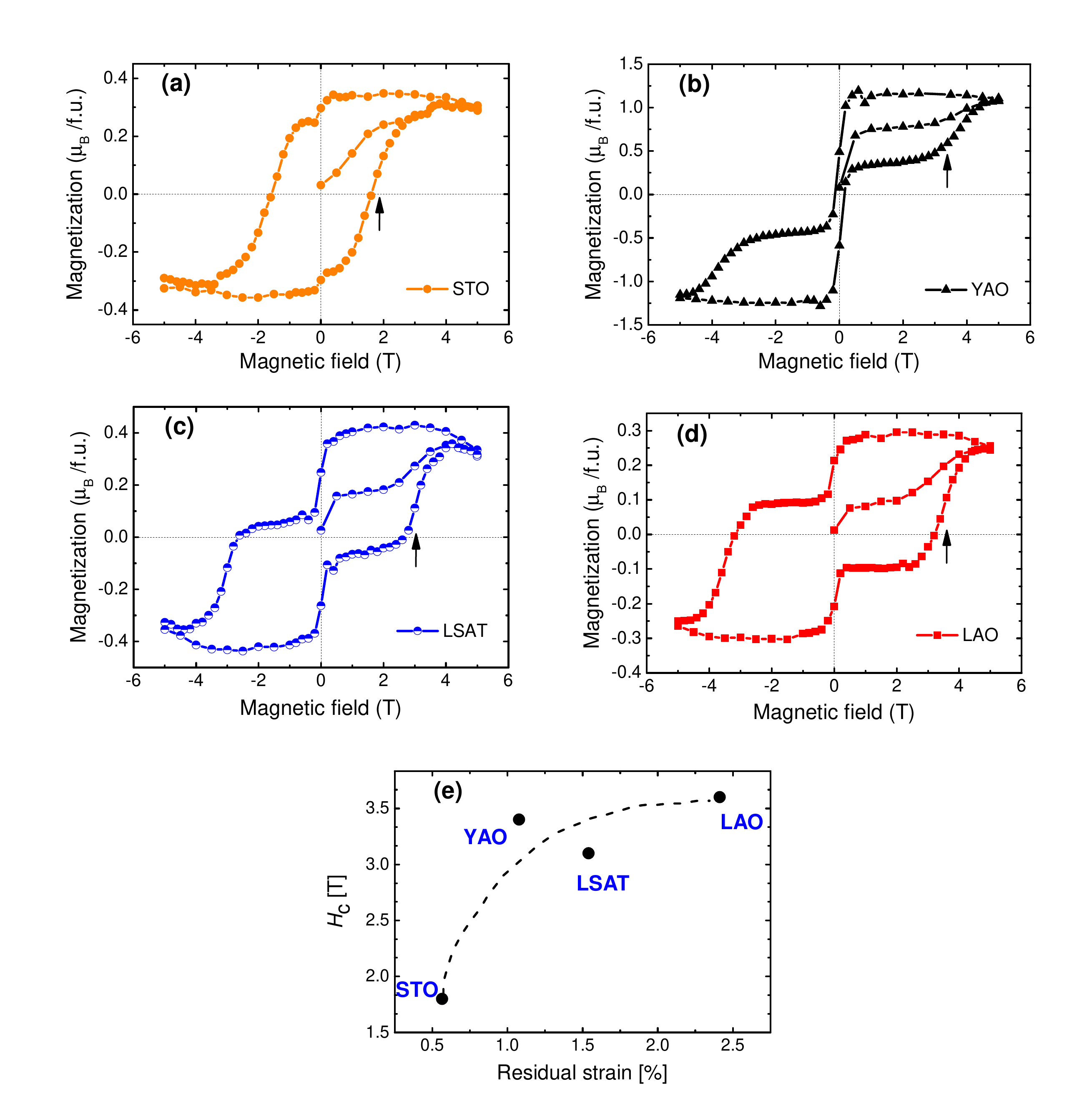}
\caption{{Magnetic measurement as a function of {\em in-plane} applied
    magnetic field for PVO films grown on :(a) STO, (b) YAO, (c) LSAT and
    (d) LAO substrate performed at 20 K. The arrow in the above figures
    represents the hard ferromagnetic component, based on which the
    coercive field was calculated. (e) Evolution of coercive
    field with residual strain. The dashed line is only guide to the eyes.}}
\label{figMH}
\end{figure}

\begin{figure}
\centering
\includegraphics[width=0.9\textwidth]{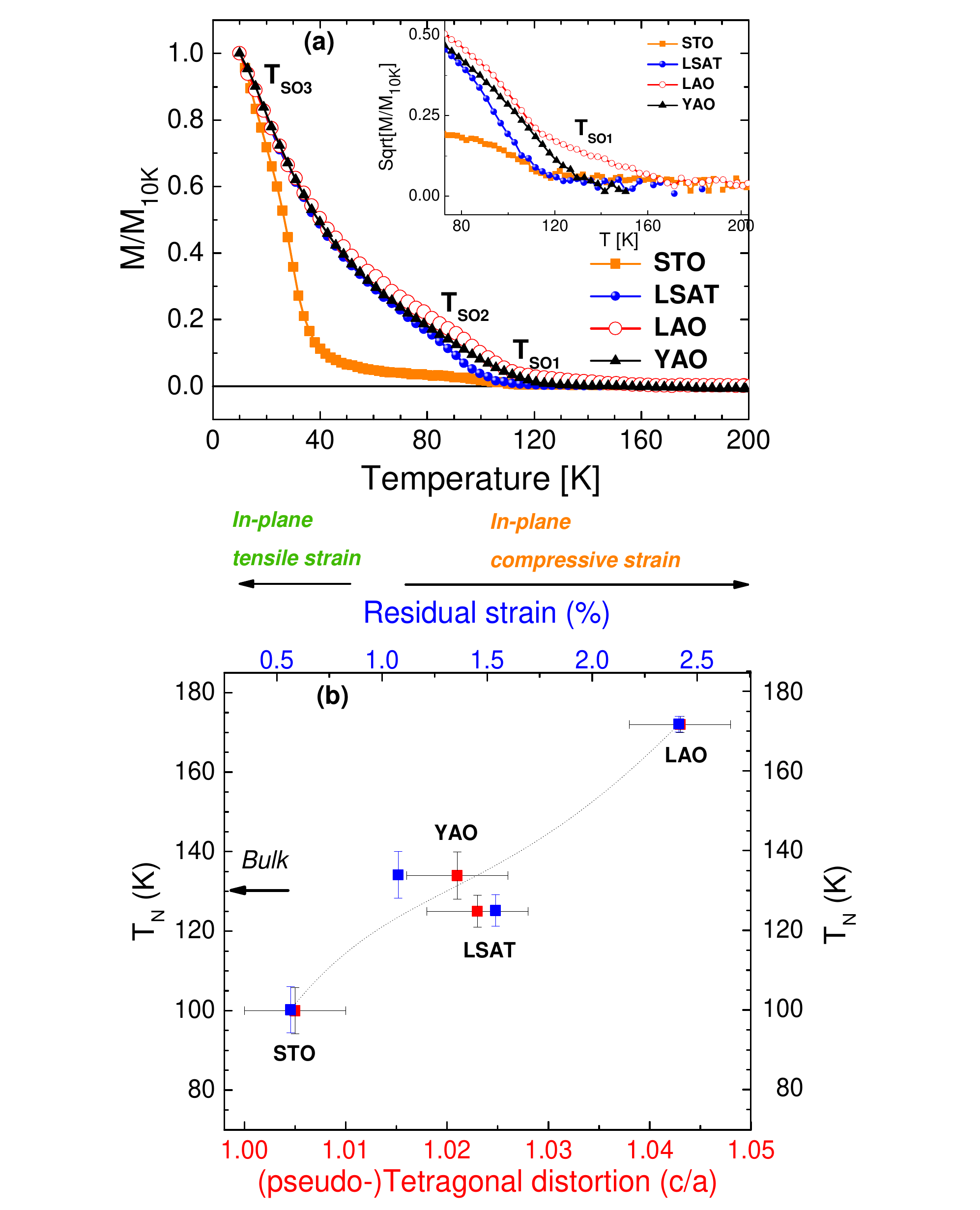}
\caption{(a) Normalized-magnetization (ratio of magnetization to the
  magnetization recorded at 10 K) dependence on temperature for PVO
  films on the YAO, LAO, LSAT and STO substrates performed at an
  \textit{in-plane} applied magnetic field of 50 \textit{Oe} after
  field-cooled at 5000 \textit{Oe}. Inset shows the magnified view of
  MT near \textit{T}$_{SO1}$. (b) : Variation of the N\'eel temperature of PVO films with
  absolute residual strain (top scale) and pseudo-tetragonal
  distortion (ratio of \textit{out-of-plane} to \textit{in-plane}
  lattice parameters, c/a)(bottom scale). The closed blue and red
  squares correspond to the \textit{T}$_N$ plot as a function of
  actual strain and tetragonal distortion respectively, along with the
  error bars. The \textit{T}$_N$ of the bulk PVO is represented by
  arrow. The dotted line is guide to the eyes.}
\label{FigExp}
\end{figure}

To examine the effect of the biaxial strain on the magnetic
properties of PVO films, the magnetization ({\em M}) of PVO films was
measured as a function of the \textit{in-plane} applied magnetic field
(\textit{H}) and temperature (\textit{T}) (Figure \ref{figMH}, \ref{FigExp}).

At low {\em T}, all PVO films show a small magnetization with a hysteresis loop indicating two magnetic phases, a soft and a hard one (figure \ref{figMH}a--d). For instance, for PVO/STO, the soft contribution shows a coercive field {\em H$_c$} at $\sim$ 0.2 T and the hard one at $\sim$ 1.8 T. While, {\em O. Copie et. al.} already observed a soft ferromagnetic behavior for the bulk PVO (our case) with {\em H$_c$} $\sim$ 0.019 T \cite{chemical strain engineering}, a hard ferromagnetic behavior was also reported for bulk PVO in Ref. \cite{Tung,bulk}. This discrepancy of coercivity between bulk and PVO
films could be explained by the microstructure. The presence of different variants of the
PVO orthorhombic cell (see TEM section) induces different pinning centers, and thus
increases the energy to return the magnetization, similar to what is observed in the orthoferrite YFeO$_3$ \cite{YFO}. Furthermore, it is interesting to note that the weightage of soft and hard magnetic phases can be modified by interplaying the epitaxial strain. 
For instance, hard and soft components were evaluated as: {\em $\Delta$M$_{hard}$} = 85 \% and {\em $\Delta$M$_{soft}$} = 15 \% for PVO/STO, {\em $\Delta$M$_{hard}$} = 70 \% and {\em $\Delta$M$_{soft}$} = 30 \% for PVO/LAO, {\em $\Delta$M$_{hard}$} = 60 \% and {\em $\Delta$M$_{soft}$} = 40 \% for PVO/LSAT, {\em $\Delta$M$_{hard}$} = 45 \% and {\em $\Delta$M$_{soft}$} = 55 \% for PVO/YAO (figure \ref{figMH}a-d), as in Ref. \cite{YFO} (Section 5 in supporting information). 
In addition, the fact that PVO/YAO film has a large percentage of soft component is interpreted as a behavior similar to the bulk PVO, since the film is fully relaxed (as shown by XRD and TEM measurements) and a small hard component might come from pinning centers due to the microstructure. Figure \ref{figMH}e shows variation of the coercivity ({\em H$_c$}) of hard magnetic phase as a function of the residual strain, whereas {\em H$_c$} of soft phase remains constant at $\sim$ 0.2 T for all substrates. 
The coercivity of hard phase changes from 1.8 T for less strained
PVO/STO film, to 3.6 T for PVO/LAO (figure \ref{figMH}e). This is presumably due to an increase of domain walls
pinning strength in more strained films.


To understand the presence of soft and hard magnetic phases versus strain, the {\em M-H} measurements were performed at
different temperatures {\em i.e} from 10 K
to 100 K. Interestingly, it was observed that the soft component is present
only at temperatures \textit{T}< 20 K for PVO/STO and PVO/LSAT but persists up to $\sim$ 80-90 K for PVO/LAO (see section 5 of supporting information). This indicates the sensitivity of the soft phase for epitaxial strain and temperature and suggests a possible magnetic ordering in PVO films
around these temperatures which triggers the rise of soft component, and will be discussed below.

In order to further investigate the effect of strain on the magnetization of PVO films, the Field
Cooled (FC) and Zero Field Cooled (ZFC) measurements were performed at
an \textit{in-plane} applied magnetic field of
\textit{H}$_{\textit{in-plane}}$ = 50 \textit{Oe}. For clarity, only
FC measurements are shown in this report with a magnified view
near \textit{T}$_N$ (or {\em T$_{SO1}$}) (Figure \ref{FigExp}a). The derivative was calculated to
visualize the magnetic transitions (see supporting information) and results are reported in Table 1.

Clearly, a magnetic transition (\textit{T}$_{SO1}$) is observed for
all the films with transition temperature ranging from 100 K for
PVO/STO, to 172 K for PVO/LAO (inset of Figure \ref{FigExp}a). This corresponds to
the magnetic transition from paramagnetic (PM) state to an antiferromagnetic (AFM) phase transition. While for bulk PrVO$_3$, the transition at {\em T$_{SO1}$} was previously ascribed to the onset of a C-type spin ordering (C-SO) of the canted vanadium moments \cite{bulkPVO,bulk}, for epitaxial PrVO$_3$ thin films, the substrate-induced strain results in a G-type SO \cite{chemical strain engineering}. The AFM N\'eel temperature (named \textit{T}$_{SO1}$ here) for PVO/STO is however
different from our previous report, where \textit{T}$_N$ $\sim$ 80 K
was reported \cite{PVO on STO}. This discrepancy could be explained by
different growth conditions (especially $\textit{P}_{O_2}$ =
10$^{-5}$ mbar, \textit{out-of-plane} lattice parameter = 3.97 \AA)
which were adopted during deposition. More interesting is
the remarkable difference of $\sim$ 70 K for the \textit{T}$_N$ of
PVO/LAO compared to PVO/STO. Notably, the {\em MT} curve also shows two
other magnetic features at \textit{T}$_{SO2}$ and \textit{T}$_{SO3}$ for LAO, YAO and LSAT,
while the former transition is strongly reduced for STO
(\textit{T}$_{SO2}$ $\sim$ 30 K). These magnetic orderings were absent
in bulk PrVO$_3$ \cite{spin-orbital}, but reported in other
orthovanadates of smaller \textit{R} ionic radii, with decreasing
temperature \cite{TmVO3,LnVO3,DyVO3}. In addition, Reehuis {\em et. al.} clearly distinguished these transitions for a doped Pr$_{1-x}$Ca$_x$VO$_3$ compound \cite{PrCaVO3}. Upon decreasing the
temperature to \textit{T}$_{SO3}$, a slight decrease in magnetization
takes place and there is a change in the slope of the magnetizations
as well as an anomaly in the inverse susceptibility (section 5 of supporting information). This
is ascribed to the FM ordering of praseodymium sublattice and/or an AFM coupling between
Pr$^{3+}$ 4\textit{f} and V$^{3+}$ 3\textit{d} moments, which results in
decrease in the net magnetization below \textit{T}$_{SO3}$. Therefore, by comparing the {\em MH} measurements
performed at different tempeatures (10 - 100 K) where a soft component
in {\em MH} was observed at temperature {\em T} $\leq$ 20 K for PVO/STO and PVO/LSAT and up to 80 K for PVO/LAO (see
supporting information) and the magnetic transition \textit{T}$_{SO3}$
in {\em MT}, we propose that the soft component in {\em MH} results from the AFM coupling between Pr and V$^{3+}$ sublattice. It is worth noting that another Pr$^{3+}$ magnetic state may exist at the surface of the PVO film. Indeed, as it has been shown in DyTiO$_3$ thin films that over-oxidation at the surface could favor a higher valence state of the transition metal oxide \cite{dead layer}. As a consequence, it would favor V$^{4+}$ and then alter the exchange interactions with Pr ions, resulting in isolated paramagnetic Pr$^{3+}$. Since the measured saturation magnetization remains low compared to 3.57 {$\mu$}$_B$ expected for isolated Pr$^{3+}$, it seems that over-oxidized surface contribution is rather small. However, the fact that the soft component contribution is modified by changing the substrate indicates rather a modification through the entire film and not only at the surface. 


Also, similar to earlier reported for bulk PrVO$_3$ \cite{PrCaVO3}, the praseodymium sublattice begins to get polarized due to presence of exchange field produced by the vanadium sublattice, resulting in a ferrimagnetic structure upon cooling. Here, a small hump at {\em T} $\sim$ 90 K is also seen, which could be the emergence of another type of spin configuration and/or a phase coexistence between C-SO and G-SO. This seems consistent with the modification of hysteresis loop as the temperature is lowered through \textit{T}$_{SO2}$, due to switching of spins or change in the spin configuration (see section 5 of supporting information). However, the feature may be also just related to the overlap of transition regime {\em T$_{SO1}$} and {\em T$_{SO3}$}. Further magnetic analysis will be published elsewhere.\par

To understand further the relationship between the magnetic properties
and strain or distortion (ratio of \textit{out-of-plane} to
\textit{in-plane} lattice parameters), the \textit{T}$_N$ versus
lattice strain is plotted for the PVO films, as shown in Figure \ref{FigExp}b. The
\textit{T}$_N$ (\textit{T}$_{SO1}$) of the PVO films increases
altogether with the residual strain, which is highest for LAO
(\textit{T}$_N$ $\sim$ 172 K) and lowest for STO (\textit{T}$_N$
$\sim$ 100 K). While PVO film on YAO has \textit{T}$_N$ close to the
bulk PVO (\textit{T}$_N$ $\sim$ 130 K). Furthermore, the PVO/YAO film
is \textit{in-plane} fully relaxed while \textit{out-of-plane} lattice
parameter is larger than the bulk. This produces a distorted structure
with $c/a$ ratio $\sim$ 1.02. The enhancement of \textit{out-of-plane}
lattice parameter of PVO/YAO might be a result of defects in film such
as oxygen vacancies etc. It is interesting to note that the influence
of small compressive strain (LSAT) in PVO film is similar to bulk, where
a small tensile strain (STO) decreases the \textit{T}$_N$ by 30 K
\cite{spin-orbital}. On the other hand, it requires a large compressive
strain of 2.4 \% (LAO) to increase the same by 40 K \textit{cf.}
bulk. \par
To further explore the magnetic properties of PVO films and their
dependence on strain, which lead to a tilting of BO$_6$
octahedra or change in the B-O-B bond angle
\cite{tilts,LVO,rotation,quantifying octahedral rotation,misfit strain
  accommodation}, it is necessary to have a complete knowledge of
distortion of the structure and the VO$_6$-octahedral rotation. From
previous studies of strained oxide perovskites, the degree of rotation
of BO$_6$ octahedra depends strongly on sign and the magnitude of the
strain \cite{quantifying octahedral rotation,misfit strain
  accommodation}.
Under tensile \textit{in-plane} strain (\textit{c/a} \textless 1.01),
the VO$_6$ octahedra comprise of an enhanced \textit{in-plane} V-O
bond length and V-O-V bond angle close to 180$^{\circ}$. This
decreases the \textit{in-plane} AFM superexchange interaction between
nearest neighbour sites, hence reduced \textit{T}$_N$. On the other
hand, under compressive \textit{in-plane} strain (\textit{c/a}
\textgreater 1.01), it is the other way round i.e. a reduced
\textit{in-plane} V-O bond length and V-O-V bond angle \textless
180$^{\circ}$. This, as a result, enhances the \textit{in-plane} AFM
interaction and therefore enhanced \textit{T}$_N$.

\section{First-principles simulations}
\begin{figure}
\centering
\includegraphics[width=1\textwidth]{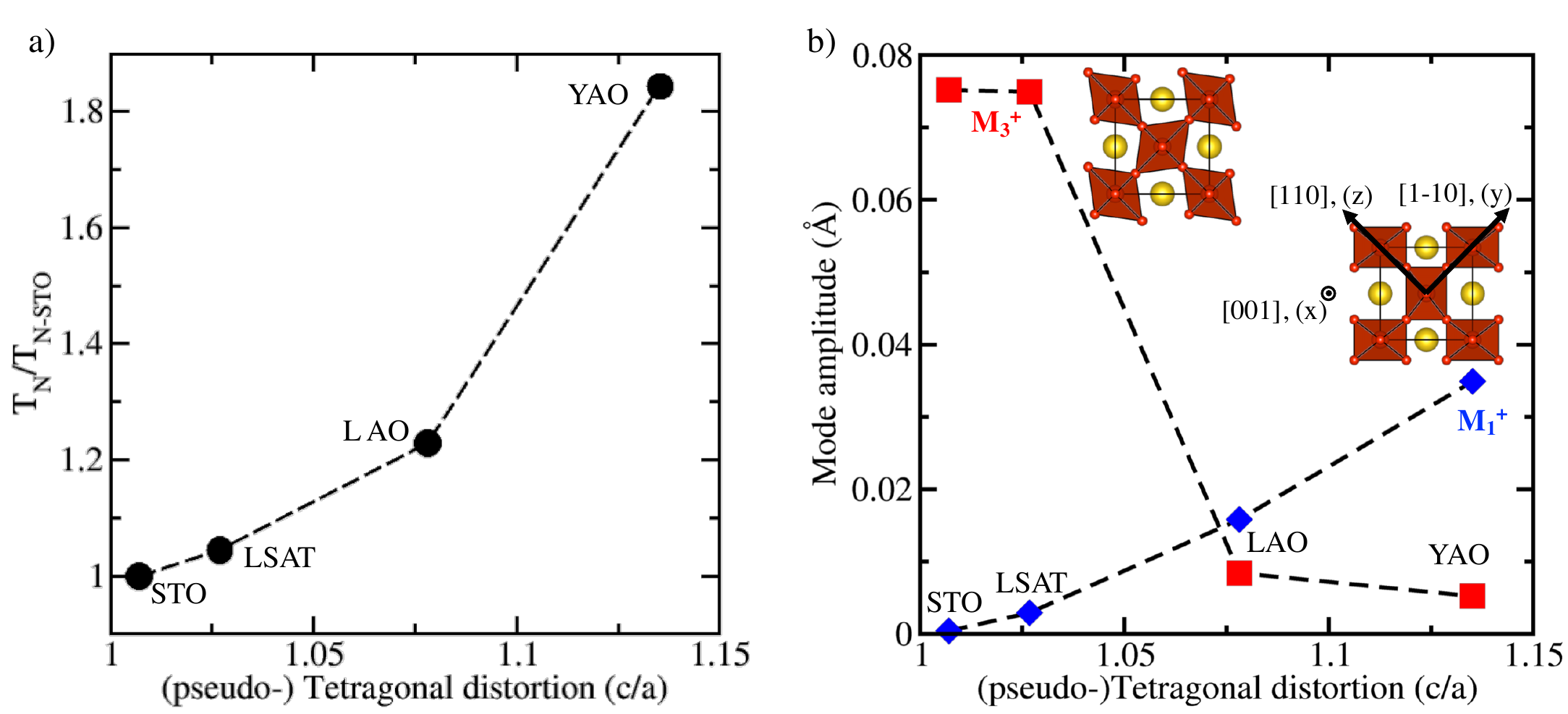}
\caption{Relative evolution of the N\'eel temperature (a) and of the
  amplitude of the M$_3^+$ and M$_1^+$ distortions as a function of
  the pseudo-tetragonality of the PrVO$_3$ films. Sketches of the
  structural distortions and local axes are presented as inserts. }
\label{figDFT}
\end{figure}

To get further insights on the role of the epitaxial strain on the
magnetic properties of PrVO$_3$ films, we have performed {\em
  first-principles} simulations using Density Functional Theory
(DFT). Consistently with previous studies \cite{chemical strain engineering},
 DFT correctly predicts that bulk PrVO$_3$
is a C-SO insulator in the ground state. Regarding the thin films, we
find that the perovskite grows with the (001) and (1-10) $Pbnm$ axes
aligned along the substrate for all the tested films ({\em e.g.}
PrVO$_3$ grown on STO, LSAT, LAO and YAO substrates, see insets of
figure \ref{figDFT}b for sketches of local axes and growth
orientation). This yields films grown along the orthorhombic (110)
direction, in sharp agreement with experiments. We emphasize here that
due to the presence of small domains in the films inducing a
mechanical constraint~\cite{chemical strain engineering}, we have considered growth
conditions with the (110)$_o$ direction forced to be orthogonal to the
substrate ({\em i.e.} the film is not allowed to tilt). With that
additional constraint, the ground state is associated with a $P2_1/m$
symmetry with nearest neighbor V$^{3+}$ spins antiferromagnetically
coupled in all crystallographic directions. It yields a G-type spin
ordering compatible with experiments. Finally, all films are
insulating with band gaps ranging from 1.50 eV (YAO) to 1.78 eV (STO).

Although mean-field methods such as DFT can not provide accurate
values of the N\'eel temperature, they nevertheless remain valuable
technics for capturing trends as a function of external
stimuli~\cite{Green function}. We report on Figure \ref{figDFT}a the
ratio of the N\'eel temperature with respect to that of PrVO$_3$ grown
on a STO substrate as a function of the pseudo tetragonality $c/a$ of
the films extracted from our simulations (see method for details on
evaluation of the N\'eel temperature). As one can see, DFT captures
the trend observed experimentally with an enhancement of the N\'eel
temperature going from STO to YAO substrates, although our computed
$T_N/T_{N-STO}$ ratio is smaller than the experimental one for the LAO
substrate. Amazingly, if the material could be stabilized on YAO
without relaxation of the film, the N\'eel temperature is expected to
be approximately multiplied by two with respect to that of PrVO$_3$
films deposited on STO.

Along with validating the experimentally measured trend for T$_N$ as a
function of the applied epitaxial strain, our {\em first-principles}
simulations also provide microscopic insights on the origin of this
physical behavior. We observe that both magnetic constants J$_1$ and
J$_2$ between nearest V$^{3+}$ neighbors along the (1-10) (or (110))
and (001) directions, respectively, increase with enlarging the
compressive epitaxial strain. Firstly, this is ascribed to shorter V-O
bond lengths along the (1-10) and (001) directions induced by
strain. Secondly, we do not observe any significant modifications of
oxygen cage rotations amplitude in the different films -- the
$a^-a^-c^0$ rotation amplitude is even slightly increasing with
decreasing the substrate lattice parameter! -- and thus the classical
``$\widehat{V-O-V}$ angles going to 180$^{\circ}$'' argument cannot
explain the strengthening of Js. Nevertheless, we find a crossover
between two lattice distortions as a function of the epitaxial strain
(see figure \ref{figDFT}b): (i) for a moderate lattice mismatch ({\em
  e.g} STO and LSAT), we extract a large Jahn-Teller distortion,
labelled M$_3^+$, producing an asymmetry of V-O bonds on nearest V
sites that is reminiscent of bulk RVO$_3$ physics and (ii) for a large
lattice mismatch ({\em e.g} LAO and YAO), the JT motion vanishes and
is replaced by a M$_1^+$ distortion unaffecting V-O bond lengths but
distorting O-V-O angles in VO$_2$ planes orthogonal to the (001)
direction (see insets of figure \ref{figDFT}b for sketches of the
distortions). The amplitude of the latter distortion, absent in the
bulk and roughly zero for films grown on STO and LSAT substrates,
closely behaves like T$_N$ as a function of the tetragonality of the
material. In fact, the crossover between the amplitude associated with
the M$_1^+$ and M$_3^+$ distortions highlights a clear modification of
the electronic structure: the two V$^{3+}$ $d$ electrons are located
in the $d_{yz}$ orbital plus an alternating combination of the
$d_{xz}\pm d_{xy}$ orbitals on neighboring sites for moderate strains
while they lie in the $d_{xz}$ and $d_{yz}$ orbitals on all
neighboring sites for large compressive epitaxial strain (see insets
of Fig\ref{figDFT}.b for the definition of local axes). It follows
that V-O bond length contractions combined with the modifications of
V$^{3+}$ $d$ orbital occupancies for LAO and YAO substrates favor
superexchange in the three crystallographic directions and thus
strongly promotes the enhancement of the N\'eel temperature.
\par

This illustrates that not only cooperative octahedral-site rotation {\em i.e} rigid octahedra tilts and rotations may tune the physical properties but also octahedral-site disortion through electronic state modifications as reported for bulk orthorhombic perovskite \cite{Intrinsic,RCrO3}. We show here that octahedral-site disortion can be driven by mechanical strain engineering and should be considered for other epitaxial orthorhombic perovskite thin films.

\section{Conclusion}

In conclusion, we have successfully grown single-phased PrVO$_3$ thin
films on top of various single crystal substrates. The most distorted structure with {\em c/a} $\sim$ 1.04 is observed on LAO substrate where a large strain of 2.4 \% is measured. Furthermore, a relationship between the magnetic properties and
the structural distortion (\textit{c/a}) in PrVO$_3$ films was
developed. We have also evidenced a clear ferromagnetic behavior of
PrVO$_3$ thin films at low temperature, and shown that the
{\em MH} hysteresis loop comprises of two magnetic sublattices, which gives
rise to a soft and a hard ferromagnetic-like component in
{\em MH}. The magnetic phase diagram (\textit{T$_N$ vs. c/a})
for PrVO$_3$ films was mapped out for 1 \textless \textit{c/a}
\textless 1.04. The most distorted film has \textit{T}$_N$ $\sim$ 172
K, 40 K higher than the bulk. Whereas, the least distorted film
has \textit{T}$_N$ $\sim$ 100 K, 30 K lower than bulk, making
PVO films an eligible candidate for application point of view for wide
range tuning of its magnetic transition temperature. Finally, the {\em first-principles} simulations have confirmed that the compressive strain not only produces stronger magnetic interactions, but also promotes electronic states totally absent of the bulk.\\

\section{acknowledgements}
The authors thank F Veillon for his valuable
experimental support. DK received his PhD from Region Normandie. This work is supported by Region Normandie,partly by french ANR POLYNASH and Labex EMC3. CUJ thanks universit\'e de Caen Normandie for his
visiting fellowship. B. Domeng$\grave{e}$s acknowledges the financial support of the program EQUIPEX GENESIS, Agence Nationale de la Recherche (ANR-11-EQPX-0020) for TEM lamella preparation. Ph. Ghosez acknowledges the F.R.S/F.N.R.S PDR
project HiT4FiT and ARC project AIMED. {\em First--principles}
calculations were performed at Abel supercomputers through the PRACE
project TheoMoMuLaM and at the Cartesius supercomputer through the
PRACE Project TheDeNoMO. The authors also took advantage of the
C\'eci--HPC facilities funded by F.R.S.--FNRS (Grant No 2.5020.1) and
the Tier--1 supercomputer of the F\'ed\'eration Wallonie--Bruxelles
funded by the Walloon Region (Grant No 1117545).

\end{document}